\documentclass[epj,twocolumn]{webofc}
\usepackage[varg]{txfonts}   

\woctitle{VI International Conference FUSION14}
\begin{document}
\title{Effect of pairing on transfer and fusion reactions
}

\author{Guillaume Scamps\inst{1}\fnsep\thanks{\email{scamps@ganil.fr}} \and
        Denis Lacroix\inst{2}\fnsep\thanks{\email{lacroix@ipno.in2p3.fr}} 
}

\institute{GANIL, CEA/DSM and CNRS/IN2P3, Boite Postale 55027, 14076 Caen Cedex, France
\and
           Institut de Physique Nucleaire, IN2P3-CNRS, Universite Paris-Sud, F-91406 Orsay Cedex, France
          }

\abstract{ In the present contribution, the effect of pairing on nuclear transfer and fusion  reactions close to the Coulomb barrier 
is discussed. A Time-Dependent Hartree-Fock + BCS (TDHF+BCS) microscopic theory has been developed 
to incorporate pairing. One- and two-particle transfer probabilities can be obtained showing the importance of pairing. The calculated transfer
probabilities are compared to the recent experimental results obtained for the $^{96}$Zr+$^{40}$Ca. Reactions involving the $^{18}$O with lead isotopes are also 
presented, that are also of current experimental interest.
Finally, a study of the fusion barrier height predicted with the TDHF+BCS theory is compared to the experimental values for the $^{40,44,48}$Ca+$^{40}$Ca reactions.
}
\maketitle
\section{Introduction}
\label{intro}

Pairing correlations are known to play an important role in the structure of the nucleus. 
It is nowadays an important challenge to take into account for the pairing correlations in state of the art microscopic transport theories \cite{Ave08,Eba10,Ste11,Has07}. 
Recently, it was shown that pairing has a non negligible effect on the description of giant resonances 
\cite{Sca13,Sca14} in particular to provide realistic properties (deformation, single-particle state fragmentation around the Fermi energy, $\dots$) 
for nuclear ground state on top of which collective excitation can be built up. These properties are also expected 
to be important to describe nuclear collisions. 
The goal of this study is to illustrate how pairing influences the transfer and fusion processes.

The natural way to incorporate pairing into a mean field dynamic, is to extend Time-Dependent Hartree-Fock theory by introducing a quasi-particle picture.
This leads to the so-called  Time-dependent Hartree-Fock-Bogoliubov (TDHFB) approach. This approach, while formally very attractive,  still requires too 
much numerical effort to be applied on a large scale \cite{Ave08,Ste11}. A good compromise able to grasps most aspects of pairing correlations while keeping 
the numerical simulation reasonable, is to consider its simplified TDHF+BCS limit \cite{Eba10,Sca12}. Recently, this approach has been applied to nuclear collision
at energy close or below the Coulomb barrier. In the present work, new comparisons with recent experiments are made. 
All details of the calculations are given in the Ref. \cite{Sca13}. 
In the section 2, we study the transfer reaction and in the section 3, the influence of pairing on the fusion 
barrier is discussed.
\begin{figure}[!ht]
\centering
\includegraphics[width= 0.9 \linewidth]{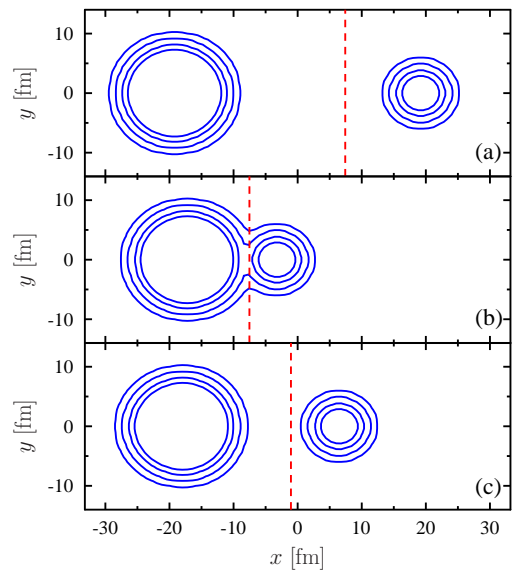}
\caption{ Evolution of the total density projected on the $x$ reaction plane for the central collision $^{208}$Pb+$^{18}$O. The center of mass energy is 71.88 MeV. The panels (a), (b) and (c) correspond respectively to the time t=0 s, t=13.5$\times$10$^{-22}$ s and t=27$\times$10$^{-22}$ s. The neck position is indicated by the dashed vertical line in each panel.}
\label{fig:film_proc}      
\end{figure}

\section{Transfer Reaction}

One of the motivation of our study is the recent renewal of interest in transfer reactions below the Coulomb barrier, like the reaction $^{96}$Zr+$^{40}$Ca \cite{Cor11}. With the specific detectors setup used in this experiment, a rather clean extraction of multi-nucleon transfer probabilities have been obtained. 
In particular, a strong enhancement of the two-particle transfer probabilities was found with $P_2\simeq3P_1^2$. Here $P_2$ and $P_1$ are respectively 
the one- and two-neutrons transfer. In the absence of correlation, quantum and fermionic effects, one simply expects a sequential transfer 
with $P_2\simeq P_1^2$. In order to understand how pairing can lead to an enhancement of $P_2$, the TDHF+BCS theory has been applied to this reaction.

In our microscopic model, the two nuclei are first separately initialized in their ground states with the Hartree-Fock plus BCS theory with the Skyrme functional Sly4d and a mixed contact type pairing interaction \cite{Ber06}. 
Then the two nuclei are positioned on a lattice with given initial velocities. The position and initial velocities are chosen to properly describe a given 
beam energy $E_B$ and impact parameter $b$. One of the advantage of the experiment  \cite{Cor11} is that it can be simulated using zero impact parameters  only, 
that greatly simplifies the theoretical description (see discussion below).  An example of density evolution below the Coulomb barrier for
a central reaction $^{208}$Pb+$^{18}$O is presented in Fig. \ref{fig:film_proc}. At this energy, the two nuclei approaches (a), can exchange particles during the contact (b) and re-separates (c).

\subsection{The $^{96}$Zr+$^{40}$Ca reaction}

The transfer probabilities of  Ref. \cite{Cor11} have been presented  in terms of the minimal distance of approach $D$. Assuming a Rutherford trajectory, a simple relation between the center of mass energy $E_{\rm c.m.}$ and $D$ is obtained,
\begin{align}
D=\frac{Z_P Z_T e^2}{2 E_{\rm c.m.}}\left( 1 + \frac1{\sin(\theta_{cm}/2)} \right)
\end{align}
with $Z_P$ and $Z_T$  the target and projectile proton numbers while $\theta_{cm}$ is the center of mass scattering angle. If we assume that the transfer probabilities depend only on the distance of closest approach, a comparison between theory and experiment can be made by performing 
several TDHF+BCS simulation where the beam energy is varied.  

\begin{figure}[!ht]
\centering
\includegraphics[width= \linewidth]{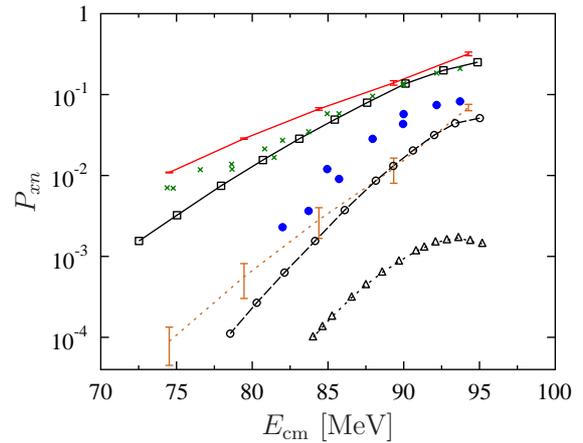}
\caption{Transfer probabilities as a function of the center of mass energy for the reaction $^{96}$Zr+$^{40}$Ca. 
The experimental data extracted from \cite{Cor11} are represented by full circles for $P_2$ and crosses for $P_1$. The TDHF+BCS results are displayed by red solid line for the probability to transfer one neutron and by orange dashed line for the probability to transfer two neutrons. 
In this figure, we also report the theoretical results obtained in Ref.   \cite{Cor11} where a shell model calculation + semi-classical treatment of the reactions have been used. The results are respectively shown by squares for the transfer of one neutron, triangles for the probability to transfer two neutrons if the only retained channel is the ground state to the ground state transfer. The open circles correspond to the two neutrons transfer where not only the GS to GS but also GS to first excited state transfer is accounted for. }
\label{fig:comp_corradi_theo}      
\end{figure}

To extract the transfer probabilities from TDHF+BCS the  double projection technique, that project out onto given neutron/proton numbers in the quasi-target and/or quasi-projectile after the reaction is used \cite{Sca13b}. The results are shown in Fig. \ref{fig:comp_corradi_theo} for the reaction $^{96}$Zr+$^{40}$Ca and compared to the experimental data as well as to the theoretical calculations presented in Ref.  \cite{Cor11}. 
As we can see, the one-particle transfer probability is slightly overestimated in our calculation. 
As mentioned in the Ref. \cite{Cor11}, the probability to transfer one neutron depends of the single particle energies and of the fragmentation of the occupation numbers. Here, no parameter are adjusted to experimental data in our calculations and the difference in $P_{1n}$ certainly stems from  the location of single-particle levels around the Fermi energy obtained  in the mean-field approach that differs from the one used in Ref. \cite{Cor11}.
\begin{figure}[!ht]
\centering
\includegraphics[width= 0.9\linewidth,clip]{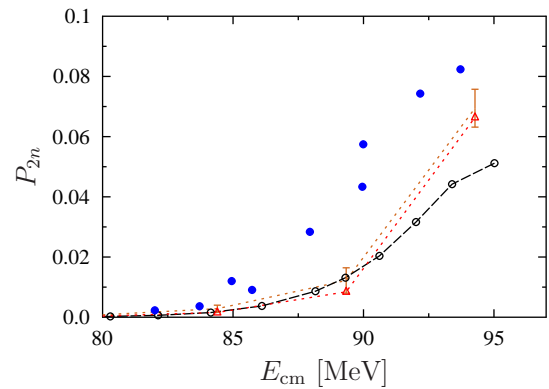}
\caption{Same as Fig. \ref{fig:comp_corradi_theo} in linear scale for the two neutrons transfer probability. The results of $P_{2n}$ obtained by neglecting the anomalous density is also shown by red triangles.  }
\label{fig:comp_corradi_theo_linear}      
\end{figure}

The situation is different for the probability to transfer two neutrons. The results found by the TDHF+BCS theory is below the experimental data by a factor 4 at 90 MeV while only 15 \% of the probability is missed at 93 MeV. Note that the situation is  better compared to the case where the anomalous density is neglected.
For this reaction, we do not show the results of TDHF, because the $^{96}$Zr in the Hartree-Fock theory, is found to be have an octupolar deformation in its ground state.
To get a more quantitative view of the missing probabilities as well as of the difference between our results and the theoretical calculations given in ref. \cite{Cor11}, the two-particle transfer probabilities are shown in linear scale in Fig. \ref{fig:comp_corradi_theo_linear}. from this figure, we can see that:
\begin{itemize}
  \item the two theoretical calculations are globally in agreement. This is an interesting point since two completely different strategies are used. In our calculation, structure and dynamical effects are included  in a unified approach while in Ref. \cite{Cor11}, static properties are obtained using nuclear structure models including 
  pairing while the nuclear reaction part is treated separately. Nevertheless, the two approaches are consistent with each others. 
  \item We see that our approach predicts slightly higher $P_{2n}$ especially at high beam energy. It should be noted, that no selection of channels is made 
  in TDHF+BCS, in particular, we do not restrict the transfer to the ground state or to the first excited state. The enhanced $P_{2n}$ plaids in favor of transfer to states at higher excitation energy that have been omitted in Ref. \cite{Cor11}.    
  \item Pairing correlations enters both through the non-zero value of the anomalous density components and through the fragmentation of 
  single-particle states. The enhancement of two-particle transfer is mainly due to the latter effect leading to a reduction of the Pauli exclusion principle 
  during transfer. This is illustrated in Fig.  \ref{fig:comp_corradi_theo_linear} where the red triangle gives the results when the anomalous density 
  is set to zero while keeping the BCS occupation numbers fixed. The results are very close to the full case.  
  
  \item Overall, we see that including pairing is not sufficient to understand the two-particle transfer and additional effects that are not included in the quasi-particle picture seems to play an important role. Among them, we could anticipate that correlations induced by the diffusion of the intrinsic state towards complex 
  configurations and/or quantal zero point motion in collective space might plays an important role. To incorporate these effects, theories beyond the quasi-particle approaches like the time dependent density matrix theory \cite{Ass09} or the stochastic mean-field  theory \cite{Yil11}  might be able to provide suitable tools.
\end{itemize}

\section{The $^{18}$O+$^{A}$Pb reactions}

Here we study the collision between the $^{18}$O and different lead isotopes. The $^{18}$O nucleus is quite interesting because (i) such nuclei can be realistically 
used in reaction (ii) it is among the lightest nucleus that enters within the range of applicability of the Energy Density Functional approach (iii) in a simplified picture, 
it can be seen as a single Cooper pair out of an inert $^{16}$O core. We investigate below possible effect of the change of the collision partner. Using the TDHF+BCS
approach, the fusion threshold that includes possible dynamical effects can be extracted \cite{Was08}. The fusion threshold is find to be at an energy of 73.85 MeV for the $^{18}$O+$^{206}$Pb and 73.35 MeV for the $^{18}$O+$^{2068}$Pb reaction.

\begin{figure}[ht]
\centering
\includegraphics[width= \linewidth,clip]{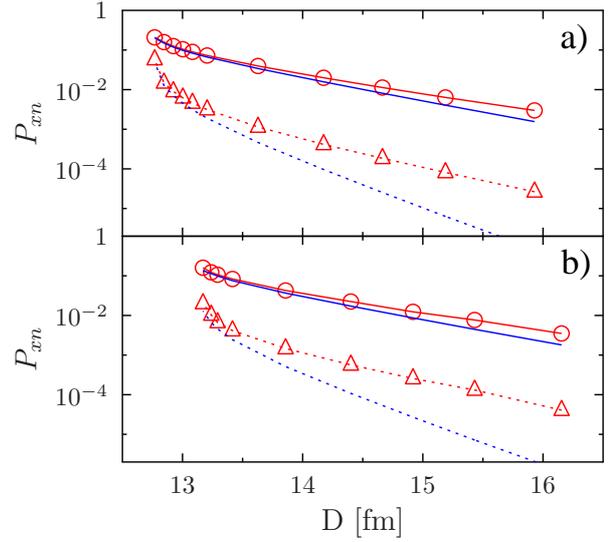}
\caption{Neutron transfer probabilities from oxygen to lead as a function of the distance of closest approach. The two reactions are shown, panel (a) : $^{206}$Pb+$^{18}$O, panel (b) : $^{208}$Pb+$^{18}$O. The results are shown for the probabilities to transfer one neutron (triangles) and two neutrons (circles). The results without pairing is also shown with solid and dashed line respectively for the probability to transfer one and two neutrons. }
\label{fig:O18Pb2xx_bcs_n}      
\end{figure}

\begin{figure}[ht]
\centering
\includegraphics[width= \linewidth,clip]{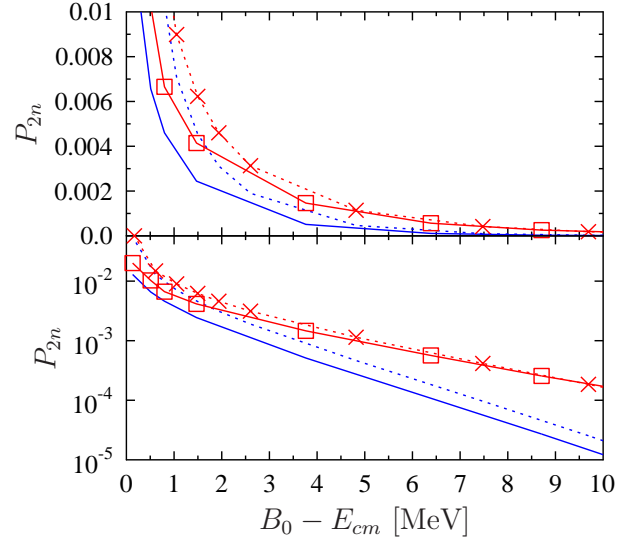}
\caption{ Two-neutron transfer probability as a function of the energy below the barrier for the reaction $^{208}$Pb+$^{18}$O (solid lines) and $^{206}$Pb+$^{18}$O (dashed lines). The results with the pairing are shown with crosses and squares and the results without pairing are shown by solid and dashed lines. }
\label{fig:O18Pb20x_P2}      
\end{figure}

The one and two-particles transfer probabilities obtained for the reactions of $^{18}$O with lead nuclei using the 
TDHF+BCS theory are shown in Fig. \ref{fig:O18Pb2xx_bcs_n} as a function of the distance of closest approach. The theory without pairing obtained with the equal-filling approximation \cite{Sca13b} is also shown.
Similarly to previous studies, a significant enhancement of the two-particle transfer is observed when pairing correlations are accounted for. 

To focus on the effect of changing the lead isotopes by 2 nucleons only, a direct comparison of the two reactions considered here is shown 
in fig. \ref{fig:O18Pb20x_P2}. We see on this figure that with the equal-filling approximation, the probability of two neutron transfer is systematically larger by a factor 2 with $^{206}$Pb than for $^{208}$Pb. A similar situation is seen also when pairing is plugged in at high energy close to the Coulomb barrier. 
The lowering of pairing in $^{208}$Pb compared to $^{206}$Pb can be anticipated from the magicity of $^{208}$Pb. This magicity has two effects (1) The 
$Q$ value associated to the reaction $2n$+$^{206}$Pb $\rightarrow$ $^{208}$Pb is higher than the one associated to $2n$+$^{208}$Pb $\rightarrow$ $^{210}$Pb, due to the $^{208}$Pb binding energy (ii) before transfer, no room is available in the $3p_{1/2}$ shell due to the sub-shell closure leading to an hindrance 
of the $2n$ transfer compared to  $^{206}$Pb. This illustration shows that the shell structure properties of the collisions partners 
should be carefully understood as well in studying  pairing effects on multi-nucleon transfer  close to the Coulomb barrier.    

\section{Systematic of fusion barrier threshold}

The role of pairing on fusion barrier has been systematically investigated using the TDHF+BCS approach by identifying 
the fusion threshold energy with the technique employed in \cite{Sim08}. 
Two effects stemming from pairing are expected to influence the barrier. First, pairing influence the ground state deformation. 
With pairing the nuclei tend to be spherical, this have an important effect on the fusion. The second effect is the transfer of neutrons before the barrier. 

In order to compare to TDHF and avoid the comparison of nuclei with different shapes, we used in that case, the equal-filling approximation that consists in filling
the last major shell of degeneracy $\Omega$ by a partial fractional occupation number $n=N/\Omega$ where $N$ is the number of nucleons to distribute in that shell. This theory is referred as the no pairing theory, then the deviation of the results between the TDHF+BCS theory and the equal-filling theory will directly 
uncover genuine effects of pairing correlations on the fusion barrier height.

\begin{figure}[ht]
\centering
\includegraphics[width= \linewidth,clip]{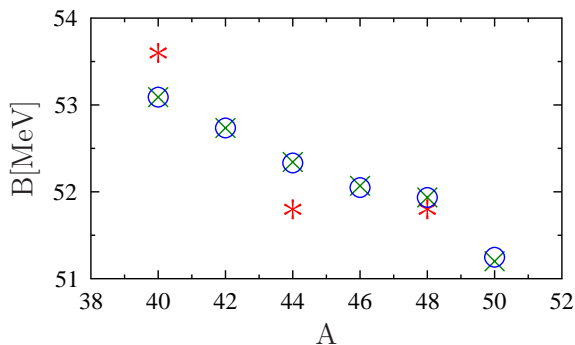}
\caption{Fusion barrier for the $^{A}$Ca+$^{40}$Ca reaction. The experimental barrier \cite{Siw04} is compared to the barrier extract from the TDHF+BCS (circles) and TDHF (crosses) calculations.}
\label{fig:fus_Ca40_Ca4X_proc}      
\end{figure}

In figure Fig. \ref{fig:fus_Ca40_Ca4X_proc}, an illustration of fusion threshold estimated with and without pairing is shown for reactions between different 
calcium isotopes. The calculation reproduces reasonably well the experimental observations. In addition, we see that pairing has almost no effect on the fusion barrier properties 

\section{Conclusion}

In the present proceedings the effect of pairing on nuclear collisions around the Coulomb barrier is investigated. Using the TDHF+BCS approach for reactions of experimental interest we show that the main effect of pairing below the Coulomb barrier is to enhance the two particles transfer while the one-particle transfer is
only slightly affected. In addition, to pairing correlations shell effects of the two collision partner can play an important role. Finally, it is shown that pairing do not affect the fusion barrier properties.

\end{document}